\documentclass[a4paper,12pt]{article}
\pdfoutput=1
\usepackage{lmodern}
\usepackage{jheppub}	
\usepackage{graphicx}				
\usepackage{amsmath}
\usepackage{mathtools}
\usepackage{amssymb}
\usepackage{booktabs}
\usepackage{comment}
\usepackage{dsfont}
\usepackage{amsthm}

\usepackage{multirow}
\usepackage{overpic}
\usepackage{caption}
\usepackage{subcaption}
\usepackage{array}
\usepackage{rotating}

\usepackage[T1]{fontenc} 

\usepackage{color}
\definecolor{dark-gray}{gray}{0.20}
\definecolor{gray}{gray}{0.30}
\definecolor{light-gray}{gray}{0.80}
\definecolor{dark-red}{rgb}{0.7,0,0}
\definecolor{dark-green}{rgb}{0.1,0.4,0}
\definecolor{dark-blue}{rgb}{0.3,0.3,0.7}
\definecolor{light-blue}{rgb}{0.8,0.8,1}

\usepackage{hyperref}
\hypersetup{
	colorlinks=true,
	linkcolor=dark-blue,
	citecolor=dark-red,
	urlcolor=dark-green,
	linktoc=page
}

\title{\boldmath Krylov Complexity of Open Quantum Systems: From Hard Spheres to Black Holes}

					\author{Vyshnav Mohan}
                                           \affiliation{Science Institute,
                                           University of Iceland \\Dunhaga 3, 107 Reykjav\'{i}k, Iceland.}
                                         \emailAdd{vyshnav.vijay.mohan@gmail.com}

\abstract{We examine the complexity of quasi-static chaotic open quantum systems. As a prototypical example, we analytically compute the Krylov complexity of a slowly leaking hard-sphere gas using Berry's conjecture. We then connect it to the holographic complexity of a $d+1$-dimensional evaporating black hole using the Complexity=Volume proposal. We model the black hole spacetime by stitching together a sequence of static Schwarzschild patches across incoming negative energy null shock waves. Under certain identification of parameters, we find the late time complexity growth rate during each quasi-static equilibrium to be the same in both systems.}

\makeatletter
\gdef\@fpheader{}
\makeatother

  {\itshape}
  {}

\begin{document} 
\newcommand{\beq}{\begin{equation}}
\newcommand{\eeq}{\end{equation}}
\newcommand{\bea}{\begin{eqnarray}}
\newcommand{\eea}{\end{eqnarray}}
\newcommand{\beas}{\begin{eqnarray*}}
\newcommand{\eeas}{\end{eqnarray*}}
\newcommand{\defi}{\stackrel{\rm def}{=}}
\newcommand{\non}{\nonumber}
\newcommand{\bquo}{\begin{quote}}
\newcommand{\enqu}{\end{quote}}
\renewcommand{\(}{\begin{equation}}
\renewcommand{\)}{\end{equation}}
\def \eqn#1#2{\begin{equation}#2\label{#1}\end{equation}}

\newtheorem*{remark}{Proposal}

\def\e{\epsilon}
\def\IZ{{\mathbb Z}}
\def\IR{{\mathbb R}}
\def\IC{{\mathbb C}}
\def\IQ{{\mathbb Q}}
\def\IH{{\mathbb H}}
\def\de{\partial}
\def\Tr{ \hbox{\rm Tr}}
\def\H{ \hbox{\rm H}}
\def\HE{ \hbox{$\rm H^{even}$}}
\def\HO{ \hbox{$\rm H^{odd}$}}
\def\K{ \hbox{\rm K}}
\def\Im{ \hbox{\rm Im}}
\def\Ker{ \hbox{\rm Ker}}
\def\const{\hbox {\rm const.}}
\def\o{\over}
\def\im{\hbox{\rm Im}}
\def\re{\hbox{\rm Re}}
\def\bra{\langle}\def\ket{\rangle}
\def\Arg{\hbox {\rm Arg}}
\def\Re{\hbox {\rm Re}}
\def\Im{\hbox {\rm Im}}
\def\exo{\hbox {\rm exp}}
\def\diag{\hbox{\rm diag}}
\def\longvert{{\rule[-2mm]{0.1mm}{7mm}}\,}
\def\a{\alpha}
\def\dag{{}^{\dagger}}
\def\tq{{\widetilde q}}
\def\p{{}^{\prime}}
\def\W{W}
\def\N{{\cal N}}
\def\hsp{,\hspace{.7cm}}

\def\br{\nonumber}
\def\IZ{{\mathbb Z}}
\def\IR{{\mathbb R}}
\def\IC{{\mathbb C}}
\def\IQ{{\mathbb Q}}
\def\IP{{\mathbb P}}
\def \eqn#1#2{\begin{equation}#2\label{#1}\end{equation}}

\newcommand{\C}{\ensuremath{\mathbb C}}
\newcommand{\Z}{\ensuremath{\mathbb Z}}
\newcommand{\R}{\ensuremath{\mathbb R}}
\newcommand{\rp}{\ensuremath{\mathbb {RP}}}
\newcommand{\cp}{\ensuremath{\mathbb {CP}}}
\newcommand{\vac}{\ensuremath{|0\rangle}}
\newcommand{\vact}{\ensuremath{|00\rangle}                    }
\newcommand{\oc}{\ensuremath{\overline{c}}}
\newcommand{\psizero}{\psi_{0}}
\newcommand{\phizero}{\phi_{0}}
\newcommand{\hzero}{h_{0}}
\newcommand{\psiin}{\psi_{\rh}}
\newcommand{\phiin}{\phi_{\rh}}
\newcommand{\hin}{h_{\rh}}
\newcommand{\rh}{r_{h}}
\newcommand{\rb}{r_{b}}
\newcommand{\psibnd}{\psi_{0}^{b}}
\newcommand{\psibndp}{\psi_{1}^{b}}
\newcommand{\phibnd}{\phi_{0}^{b}}
\newcommand{\phibndp}{\phi_{1}^{b}}
\newcommand{\gbnd}{g_{0}^{b}}
\newcommand{\hbnd}{h_{0}^{b}}
\newcommand{\zh}{z_{h}}
\newcommand{\zb}{z_{b}}
\newcommand{\man}{\mathcal{M}}
\newcommand{\hbr}{\bar{h}}
\newcommand{\tbr}{\bar{t}}

\maketitle
\flushbottom

\section{Introduction}
\label{sec:intro}
In recent years, the complexity of quantum systems has shown much promise as a tool in diagnosing quantum chaos \cite{Parker:2018yvk,Rabinovici:2022beu,Espanol:2022cqr,Hashimoto:2023swv,Camargo:2023eev}. Complexity tracks the degrees of freedom of the system, especially at very long time scales where other quantum information theoretic measures like entanglement entropy have saturated \cite{Susskind:2014moa}. This feature has been particularly fruitful in theories involving black holes, where the volume of spacelike extremal codimension-1 surfaces has been conjectured to measure the complexity of the corresponding state in the dual boundary theory \cite{Susskind:2014rva,Stanford:2014jda}. The proposal, dubbed the Complexity=Volume (CV) prescription, adds a non-trivial entry to the holographic dictionary.

In a quantum mechanical system, we can quantify complexity using \textit{Krylov Complexity} (or K-complexity) \cite{Parker:2018yvk,Barbon:2019wsy}, which measures the growth of operators by treating its evolution as a particle hopping on a semi-infinite chain\footnote{For applications of Krylov complexity to various systems, refer to \cite{Dymarsky:2021bjq,Avdoshkin:2022xuw,Camargo:2022rnt,Caputa:2021ori,Adhikari:2022whf,Kundu:2023hbk,Patramanis:2023cwz,Adhikari:2022oxr,Bhattacharjee:2022vlt,Vasli:2023syq,Rabinovici:2023yex,Jian:2020qpp,Erdmenger:2023wjg,Hornedal:2022pkc}.}. Under Hamiltonian evolution, a wavefunction centered around the first site will spread deeper into the chain. Krylov complexity is defined as the average position of the particle on the chain as a function of time, thereby capturing the ``spread'' of the operator in the operator space.

In section \ref{Kcompsection}, we will use this formulation of operator complexity to study the growth of operators in a slowly leaking hard sphere gas. We will assume that the gas is leaking from a small box into a bigger box, as in \cite{Krishnan:2021faa}. Moreover, we will work in the semiclassical, low density limit of the system. A box of hard sphere gas is classically chaotic and satisfies \textit{Berry's conjecture}, which states that the high-lying energy eigenstates behave as if they have been picked from a Gaussian ensemble \cite{berry1983semiclassical,berry1991chaos, Srednicki:1994mfb}. Berry’s conjecture played a crucial role in arriving at Eigenstate Thermalization Hypothesis (ETH) \cite{Srednicki:1994mfb} (see also \cite{PhysRevA.43.2046}) and makes the system analytically tractable.

However, since we are dealing with an open quantum system, the canonical Krylov complexity calculations would not work here. This is because the evolution is non-unitary, and the \textit{Lanczos algorithm} one usually employs to calculate the K-complexity leads to unsatisfactory
results \cite{Bhattacharya:2022gbz,Liu:2022god,Bhattacharjee:2022lzy,Bhattacharya:2023zqt}. We will sidestep this difficulty by working with a slowly leaking gas. This allows us to focus on a time period, which we will refer to as an \textit{epoch}, during which the boxes equilibrate separately, and there is no overall exchange of particles. During an epoch, the Lanczos algorithm gives meaningful results, and we can compute the Krylov complexity using standard techniques. Moreover, we will also show that we can patch together adjacent epochs to form a continuous curve.

Using ETH and very mild assumptions on the off-diagonal matrix elements of the operators, we will argue how the hard sphere calculation can be generalized to any chaotic open quantum system. This prompts us to look at our slowly leaking hard sphere gas model as an excellent prototype where analytic calculations can be carried out explicitly.

In section \ref{cvsection}, we will use the CV prescription to calculate the holographic complexity of an evaporating black hole. We will model an evaporating black hole by stitching together a sequence of static Schwarzschild spacetimes across incoming negative energy null shock waves. Each Schwarzschild patch is characterized by a constant mass that decreases as we go across the shock waves. These patches correspond to periods where the black hole is effectively not evaporating. Therefore, we will refer to them as epochs, akin to our slowly leaking gas calculation. We will calculate the volume of boundary-anchored extremal codimension-1 surfaces in this background. Under an identification of parameters, we show that the late time rate of growth of complexity during an epoch matches the slowly leaking gas calculation.

\section{Krylov Complexity of a Slowly Leaking Gas}
\label{Kcompsection}
Consider an operator $O$ that acts on the states of a quantum mechanical system. If the Hamiltonian of the system is $H$, then the time evolution of this operator is given by 
\bea
O(t)=e^{i H t} O e^{-i H t}=e^{i \mathcal{L} t} O
\eea
where $\mathcal{L}(O) = [H,O]$ is the Liouvillian superoperator. Krylov complexity measures the spread of $O(t)$ in the \textit{Krylov subspace}, the Hilbert space spanned by $\mathcal{L}^{n}O$. We will not provide a pedagogical review of Krylov complexity here as detailed reviews can be found elsewhere \cite{Barbon:2019wsy,Parker:2018yvk,Caputa:2021sib}.

The microcanonical refinement of Krylov complexity, as introduced in \cite{Kar:2021nbm}, will be our focus for analysis. Additionally, we will use the moment method to compute Krylov complexity. Let us quickly review this construction. Our starting point is the thermal two-point function of the operator:
\bea
G(t) = \sum_{i,j}e^{-\frac{\beta}{2}\left(E_i+E_j\right)}e^{it\left(E_i-E_j\right)}\left|\left\langle E_i\left|O\right| E_j\right\rangle\right|^2
\eea
Here $\beta$ is the inverse temperature, and $E_{i,j}$ are the energy eigenvalues of the system. Approximating the sum by an integral over the density of eigenstates $\rho(E)$, we get
\bea
G(t)=\int_{0}^{\infty}dEe^{-\beta E}\int_{-2 E}^{2 E} d \omega \rho\left(E+\frac{\omega}{2}\right)\rho\left(E-\frac{\omega}{2}\right) \left|\left\langle E+\frac{\omega}{2}\left|O\right| E-\frac{\omega}{2}\right\rangle\right|^2 e^{i \omega t}
\eea
where we have defined the average energy and the energy difference as follows:
\bea
E= \frac{E_i+E_j}{2} \quad \text{and} \quad \omega = E_i-E_j. \label{energyaverageeq}
\eea
The Liouvillian is sensitive only to the energy differences $\omega$, and it does not mix different average energy $E$ sectors \cite{Kar:2021nbm}. Therefore, we can work with a fixed $E$ and then average over all the other energy sectors at the end of the calculation. The fixed energy two point function can be obtained by taking its inverse Laplace transform:
\bea
G_E(t)=\int_{-2 E}^{2 E} d \omega  \rho\left(E+\frac{\omega}{2}\right)\rho\left(E-\frac{\omega}{2}\right)\left|\left\langle E+\frac{\omega}{2}\left|P_1\right| E-\frac{\omega}{2}\right\rangle\right|^2 e^{i \omega t}.
\eea
The key elements in our analysis are the moments of these two point functions. They are given by
\bea
\mu_n^E=\frac{\left.\left(-i \frac{d}{d t}\right)^n G_E(t)\right|_{t=0}}{G_E(0)} \label{momentdefeq}
\eea
Using the Hankel transformation matrix $M_{ij} = \mu_{i+j}^E$, we can immediately calculate the \textit{Lanczos coefficients} $b_n^{E}$:
\bea
\left(b_1^{E}\right)^{2 n} \left(b_2^{E}\right)^{2 n-2} \ldots \left(b_n^{E}\right)^2=\operatorname{det}\left[M_{i j}\right]_{0 \leq i, j \leq n}
\eea
The Lanczos coefficients are handy objects because they contain all the information about the dynamics of the operator $O$. Moreover, they completely determine the Krylov complexity of the operator. To see this, let us note that the fixed energy Krylov complexity is given in terms of the $K$-wavefunctions $\phi_{E,n}$ as follows \cite{Kar:2021nbm}
\bea
K^{E}(t) = \sum_{n=0}^{D_O-1}n\left|\phi_{E,n}(t)\right|^2, \label{fixedekrylovdef}
\eea
where $D_O$ is the dimensionality of the Krylov subspace. The $K$-wavefunctions are, in turn, related to the Lanczos coefficients through the Schr\"{o}dinger equation:
\bea
\dot{\phi}_{E, n}(t)=b_{n+1}^E \phi_{E, n+1}(t)-b_n^E \phi_{E, n-1}(t) .\label{kwavefunctions}
\eea
Using the initial condition $\phi_{E, n}(0)=\delta_{n0}$, we can solve for the $K$-wavefunctions and compute $K^{E}$.

Krylov Complexity $K^{E}$ measures the complexity growth of operators within a specific energy sector. Operationally, this definition of complexity corresponds to the choice of a ``microcanonical'' inner product on the Krylov subspace:
\bea
(A| B)_E=\int_{-2 E}^{2 E} d \omega \rho\left(E+\frac{\omega}{2}\right)\rho\left(E-\frac{\omega}{2}\right)  \left\langle E+\frac{\omega}{2}\left|A^{*}\right| E-\frac{\omega}{2}\right\rangle \left\langle E+\frac{\omega}{2}\left|B\right| E-\frac{\omega}{2}\right\rangle.\nonumber\\
\eea
If we had instead used a thermal inner product that involved integrating over various energy sectors, there would have been a risk of the signatures of chaos being washed out (See \cite{Kar:2021nbm} for details). Therefore, the natural thing to do is to compute the growth in a fixed energy sector and then take an average over various values of $E$. Since different energy sectors are thermally populated, we can define the \textit{thermal Krylov complexity} by taking a Laplace transform of its fixed energy counterpart \cite{Kar:2021nbm}:
\bea
K_{t h}(t) =\frac{\int_0^{\infty} d E e^{-\beta E} \mathcal{C}(E) K^{E}(t) }{\int_0^{\infty} d E e^{-\beta E} \mathcal{C}(E)}\label{thermalKrylovdef}
\eea
where the normalization constant $ \mathcal{C}(E)$ is given by 
\bea
\mathcal{C}(E)= \int_{-2 E}^{2 E} d \omega  \rho\left(E+\frac{\omega}{2}\right)\rho\left(E-\frac{\omega}{2}\right) \left|\left\langle E+\frac{\omega}{2}\left|O\right| E-\frac{\omega}{2}\right\rangle\right|^2. \label{Cnormalizationeq}
\eea
Our primary focus is on the thermal Krylov complexity. In the following subsections, we will use the moments of the thermal two point functions to calculate $K_{t h}(t)$ of a slowly leaking hard sphere gas.
\subsection{Warm-up: Single Box}
Before we look at the slowly leaking gas, it is instructive to look at a single box of hard sphere gas. Consider a cubic box of edge length $L+2a$ enclosing $N$ hard spheres. We will assume that the hard spheres are identical and have radius $a$. The classical Hamiltonian of the system is given by
\bea
H=\sum_{i=1}^N \frac{\mathbf{p}_i^2}{2 m}+\sum_{i<j} V\left(\left|\mathbf{x}_i-\mathbf{x}_j\right|\right)
\eea
where
\bea
V(r)= \begin{cases}+\infty & \text { for } r<2 a \\ 0 & \text { for } r>2 a\end{cases}
\eea
This system is classically chaotic and shows eigenstate thermalization when treated quantum mechanically \cite{Srednicki:1994mfb}. Let us denote the energy eigenfunctions of the system by $\psi(X)$, where $X$ is the $3N$-dimensinal position vector. The wavefunctions are defined on the domain
\bea
D=\left\{\mathrm{x}_1, \ldots, \mathrm{x}_N \big{|}x_{i 1,2,3} \in \left[-\frac{1}{2} L,\frac{1}{2} L\right] ;\left| \mathbf{x}_i-\mathbf{x}_j \right| \geq 2 a\right\}
\eea
We will impose the boundary condition that $\psi(X)$ vanishes on $\partial D$. The wavefunctions with energy $E_\alpha$ can be chosen to be of the following form \cite{Srednicki:1994mfb}:
\bea
\psi_\alpha(\mathbf{X})=\mathcal{N}_\alpha \int d^{3 N} P A_\alpha(\mathbf{P}) \delta\left(\mathbf{P}^2-2 m E_\alpha\right) \exp (i \mathbf{P} \cdot \mathbf{X} / \hbar)
\eea
where $\mathcal{N}_\alpha$ is the normalization constant. We can choose the wavefunction to be everywhere real by imposing $A_\alpha^*(\mathbf{P}) = A_\alpha(-\mathbf{P})$.

Let us define the thermal wavelength $\lambda$ of the system as follows:
\bea
\lambda = \sqrt{\frac{2\pi\hbar^2}{mkT_\alpha}}
\eea
where the termperature $T_\alpha$ is related to the energy through the relation $E_\alpha = \frac{3}{2}NkT_\alpha$. When the thermal wavelength $\lambda \lesssim a$, then the high-lying energy eigenstates are expected to satisfy the \textit{Berry's conjecture}, which states that $A_\alpha(\mathbf{P})$ can be regarded as a Gaussian variable with the two-point function
\bea
\left\langle A_\alpha(\mathbf{P}) A_\beta\left(\mathbf{P}^{\prime}\right)\right\rangle_{\mathrm{EE}}=\delta_{\alpha \beta} \frac{\delta^{3 N}\left(\mathbf{P}+\mathbf{P}^{\prime}\right) }{ \delta\left(\mathbf{P}^2-\mathbf{P}^{\prime 2}\right)}.
\eea
The subscript $\mathrm{EE}$ stands for \textit{eigenstate ensemble}, the fictitious Gaussian ensemble the high-lying wavefunctions can be thought to belong to. The wavefunction in the momentum space is given by:
\bea
\widetilde{\psi}_\alpha(\mathbf{P}) \equiv h^{-3 N / 2} \int_D d^{3 N} X \psi_\alpha(\mathbf{X}) \exp (-i \mathbf{P} \cdot \mathbf{X} / \hbar)
\eea

We will assume that we always work in the regime where Berry's conjecture holds ($\lambda \lesssim a$). Moreover, we will also assume that the density of the gas is low, that is, $Na \ll L^3$. These assumptions give us enormous analytic control over the system. In particular, it is easy to see that the averaged two point functions are given by \cite{Srednicki:1994mfb}
\bea
\left\langle\widetilde{\psi}_\alpha^*(\mathbf{P}) \widetilde{\psi}_\beta\left(\mathbf{P}^{\prime}\right)\right\rangle_{\mathrm{EE}}=\delta_{\alpha \beta} \mathcal{N}_\alpha^2 h^{3 N} \delta\left(\mathbf{P}^2-2 m E_\alpha\right) \delta_D^{3 N}\left(\mathbf{P}-\mathbf{P}^{\prime}\right)\label{twopointavereq}
\eea
where
\bea
\delta_D^{3 N}(\mathbf{K}) \equiv h^{-3 N} \int_D d^{3 N} X \exp (i \mathbf{K} \cdot \mathbf{X} / \hbar).\label{boxdiraceq}
\eea
\subsection{Krylov Complexity from Moments}
Consider the operator $P_1$ that measures the momentum of one of the particles. The matrix elements of this operator in the energy eigenbasis are given by
\bea
\bra E_m| P_1 |E_n \ket = \int d^{3}p_1d^{3}p_2 \cdots d^{3}p_{N} \ |p_1| \ \widetilde{\psi}^{*}_{m} (\textbf{P})\widetilde{\psi}_{n} (\textbf{P}),\label{overlapmomeq}
\eea
where $|p_1|$ is the magnitude of the momenta of one of the particles. Let us calculate the Krylov complexity of this operator by using the moments of the thermal two point function. Using \eqref{momentdefeq}, we get
\bea
\mu_{2 n}^{E} = \frac{1}{\mathcal{C}(E)}\int_{-2E}^{2E} d\omega \ \rho\left(E+\frac{\omega}{2}\right)\rho\left(E-\frac{\omega}{2}\right) \left|\left\langle E+\frac{\omega}{2}\left|P_1\right| E-\frac{\omega}{2}\right\rangle\right|^2 \omega^{2n} \label{momenteq}
\eea
$\mathcal{C}(E)$ is the normalization constant we defined in \eqref{Cnormalizationeq}. The density of eigenstates $\rho$ is given by \cite{Srednicki:1994mfb}
\bea
\rho(E)= \frac{1}{\Gamma(3N/2)E}\left(\frac{ mL^2E}{2\pi\hbar^2}\right)^{\frac{3N}{2}}.
\eea
Now let us calculate the average of the moments in the eigenstate ensemble:
\bea
\left\langle\mu_{2 n}^{E}\right\rangle_{\text{EE}} = \frac{1}{\mathcal{C}(E)}\int_{-2E}^{2E} d\omega \ \rho_0(E, \omega) \left|\left\langle E+\frac{\omega}{2}\left|P_1\right| E-\frac{\omega}{2}\right\rangle\right|^2_{\text{EE}} \omega^{2n} \label{averagedmomentexpression0}
\eea
Here we have used the shorthand notation $\rho_0(E, \omega)$ for the product of the density of states. Using \eqref{overlapmomeq}, we see that:
\bea
\left|\left\langle E_m \left|P_1\right| E_\ell\right\rangle\right|^2_{\text{EE}} = \int d^{3N}P d^{3N}P^{\prime} \ |p_1|| p^{\prime}_1| \left\langle\widetilde{\psi}^{*}_{m} (\textbf{P})\widetilde{\psi}_{\ell} (\textbf{P}) \widetilde{\psi}^{*}_{\ell} (\textbf{P}^{\prime})\widetilde{\psi}_{m} (\textbf{P}^{\prime}) \right\rangle_{\text{EE}}
\eea
The four-point function can be broken down into two point functions using Wick contractions:
\bea
\begin{aligned}
\left\langle\widetilde{\psi}^{*}_{m} (\textbf{P})\widetilde{\psi}_{\ell} (\textbf{P}) \widetilde{\psi}^{*}_{\ell} (\textbf{P}^{\prime})\widetilde{\psi}_{m} (\textbf{P}^{\prime}) \right\rangle_{\text{EE}} &= \left\langle\widetilde{\psi}^{*}_{m} (\textbf{P})\widetilde{\psi}_{\ell} (\textbf{P}) \right\rangle_{\text{EE}}\left\langle\widetilde{\psi}^{*}_{\ell} (\textbf{P}^{\prime})\widetilde{\psi}_{m} (\textbf{P}^{\prime}) \right\rangle_{\text{EE}}\\
&+ \left\langle\widetilde{\psi}^{*}_{m} (\textbf{P})\widetilde{\psi}^{*}_{\ell} (\textbf{P}^{\prime}) \right\rangle_{\text{EE}}\left\langle\widetilde{\psi}_{\ell} (\textbf{P})\widetilde{\psi}_{m} (\textbf{P}^{\prime}) \right\rangle_{\text{EE}}\\
&+\left\langle\widetilde{\psi}_{\ell} (\textbf{P})\widetilde{\psi}^{*}_{\ell} (\textbf{P}^{\prime}) \right\rangle_{\text{EE}}\left\langle\widetilde{\psi}^{*}_{m} (\textbf{P})\widetilde{\psi}_{m} (\textbf{P}^{\prime}) \right\rangle_{\text{EE}}
\end{aligned}\label{wickconteq}
\eea
From \eqref{twopointavereq}, we can see that contracting two eigenfunctions will produce a delta function in its indices. The only terms in \eqref{wickconteq} which would contribute to the moment calculation are the ones without any $\delta_{m\ell}$ factor. This is because \eqref{averagedmomentexpression0} contains a factor $(E_{\ell}-E_m)$, multpyling the four-point functions. Therefore, only the last term in \eqref{wickconteq} would contribute. Let us look at the integral of this term separately:
\bea
\Phi_{m\ell}=\int d^{3N}P d^{3N}P^{\prime} \ p_1 p^{\prime}_1 \left\langle\widetilde{\psi}_{\ell} (\textbf{P})\widetilde{\psi}^{*}_{\ell} (\textbf{P}^{\prime}) \right\rangle_{\text{EE}}\left\langle\widetilde{\psi}^{*}_{m} (\textbf{P})\widetilde{\psi}_{m} (\textbf{P}^{\prime}) \right\rangle_{\text{EE}}
\eea
Using \eqref{twopointavereq}, we see that
\bea
\Phi_{m\ell}=\mathcal{N}_m^2\mathcal{N}_\ell^2(Lh)^{3 N}\int d^{3N}P d^{3N}P^{\prime} \ p_1 p^{\prime}_1 \delta\left(\mathbf{P}^2-2 m E_{\ell}\right) \delta\left(\mathbf{P}^{\prime}{}^2-2 m E_m\right) \delta_D^{3 N}\left(\mathbf{P}-\mathbf{P}^{\prime}\right)\nonumber\\\label{operatorwavefunctioneq}
\eea
where we have used \cite{Srednicki:1994mfb}
\bea
(\delta_D^{3 N}\left(\mathbf{P}-\mathbf{P}^{\prime}\right))^2 = (L/h)^{3N}\delta_D^{3 N}\left(\mathbf{P}-\mathbf{P}^{\prime}\right).
\eea
Now let us look at the $m=\ell$ component. This corresponds to the case where $\omega=0$ and $E_m=E_\ell = E$. In the low density limit, we can essentially replace $\delta_D^{3 N}\left(\mathbf{P}-\mathbf{P}^{\prime}\right)$ with a dirac delta. This gives us
\bea
\Phi_{mm}= \mathcal{N}^4(Lh)^{3 N}\int d^{3N}P \ p_1^2 \delta\left(\mathbf{P}^2-2 m E\right)
\eea
Choosing the normalization constant as in \cite{Srednicki:1994mfb}
\bea
\mathcal{N}^{-2} \equiv \mathcal{N}_i^{-2} = L^{3N}\frac{ (2\pi mE)^{\frac{3N}{2}},}{\Gamma(3N/2)E}
\eea
we get
\bea
\begin{aligned}
\Phi_{mm}&=\mathcal{N}^2 h^{3 N}\int d^{3}p_1 \ p_1^2 \left(2 \pi m k T\right)^{-3 / 2} e^{-\mathbf{p}_1^2 / 2 m k T}\\
&=4\pi\mathcal{N}^2h^{3 N}\int dp_1 \ p_1^4 \left(2 \pi m k T\right)^{-3 / 2} e^{-\mathbf{p}_1^2 / 2 m k T}\\
& = 3 m k T \mathcal{N}^2(h)^{3 N}\\
& = (h/L)^{3 N} 3 m k T \frac{\Gamma(3N/2)(2mE)}{\left(2m\pi E\right)^{3N/2}}\\
& \equiv \Phi_E
\end{aligned}
\eea
Now let us return to the non-diagonal elements of $\Phi_{m\ell}$. From our definition \eqref{boxdiraceq}, it is easy to see that $ \delta_D^{3 N}\left(\mathbf{P}-\mathbf{P}^{\prime}\right)$ is a sharply peaked function that is zero almost everywhere. Let us choose this function to be a Gaussian distribution as in \cite{Srednicki:1994mfb}:
\bea
\delta_D^{3 N}\left(\mathbf{P}-\mathbf{P}^{\prime}\right) \simeq(L / h)^{3 N} \exp \left[-\left(\mathbf{P}-\mathbf{P}^{\prime}\right)^2 L^2 / 4 \pi \hbar^2\right]
\eea
This gives us \cite{Srednicki:1994mfb}
\bea
\Phi_{ij} \simeq \Phi_{ii} \exp \left[-m\left(E_i-E_j\right)^2 L^2 / 8 \pi \hbar^2 E_i\right]
\eea
In the notation $E_i = E+\frac{\omega}{2}$ and $E_j = E-\frac{\omega}{2}$, the expression simplifies to
\bea
\Phi_{ij} \simeq \Phi_{E} \exp \left[-\frac{m\omega^2 L^2 }{ 8 \pi \hbar^2 E}\right]
\eea
Plugging in the expressions, we find that the following integral gives the moments:
\bea
\left\langle\mu_{2 n}^{E}\right\rangle_{\text{EE}} = \frac{1}{\mathcal{C}(E)}\int_{-2E}^{2E} d\omega \ \rho_0\left(E,\omega\right) \omega^{2n} \Phi_{E} \exp \left[-\frac{m\omega^2 L^2 }{ 8 \pi \hbar^2 E}\right] \label{momentintegraleq2.1}
\eea
\subsubsection{A Tale of Two Saddles}
We will compute the moment integral \eqref{momentintegraleq2.1} using a saddle point approximation. We claim that for small moments $n$, the saddle point $\omega^{*}$ is located at
\bea
\omega^{*} \ll 2E.\label{firstsaddleconsistanteq}
\eea
To see this, let us work out the saddle point equation up to the leading order in $\omega/2E$. This gives us
\bea
\frac{2n}{\omega_{*}}-\frac{m\omega^{*} L^2}{4 \pi \hbar^2 E}=0
\eea
Solving the above equation, we find that the saddle point is located at
\bea
\omega_{*} =\sqrt{\frac{8\lambda^2E^2n}{3L^2N}} \label{saddlepointeq}
\eea
where $\lambda$ is the thermal wavelength at the energy $E$. When $n \ll \frac{3N L^2}{2\lambda^2}$, $\omega^{*}$ is much smaller than $2E$, as advertised. Substituting the saddle point in \eqref{momentintegraleq2.1}, we get:
\bea
\left\langle\mu_{2 n}^{E}\right\rangle_{\text{EE}} \simeq \frac{1}{\mathcal{C}(E)} \rho_0\left(E,\omega_{*}\right) \ \omega_{*}^{2n} \ \Phi_{E} \ \exp \left[-\frac{m\omega_{*}^2 L^2 }{ 8 \pi \hbar^2 E}\right]
\eea
When $n$ is sufficiently large, we can calculate the Lanczos coefficients using the relation \cite{Barbon:2019wsy, Parker:2018yvk,Kar:2021nbm}
\bea
\mu_{2 n} \sim\left(b_{n}\right)^{2 n}e^{o(n)}. \label{momentlanczos}
\eea
We find that
\bea
\left\langle b_n^E \right\rangle_{\text{EE}} \sim \sqrt{\frac{8\lambda^2E^2n}{3L^2N}}.
\eea
This behavior is termed ``Lanczos ascent'' in literature \cite{Kar:2021nbm}. When $n \gg \frac{3N L^2}{2\lambda^2}$, $\omega_{*} \gg 2E$. Since this violates our assumption \eqref{firstsaddleconsistanteq}, \eqref{saddlepointeq} ceases to be a saddle point of the moment integral. For large $n$, \eqref{saddlepointeq} gets replaced by a new saddle point located at $\omega_{*} \simeq 2E$. We can obtain this saddle point mechanically by noting that if\footnote{Note that the $\frac{\lambda}{L}$ is a tiny number.}
\bea
\left|\frac{\omega}{2}-E\right| \ll \frac{2\lambda^2E}{L^2}, \label{secondsaddleconsistanteq}
\eea
the saddle point equation is given by
\bea
\frac{d}{d\omega}\left( \rho_0\left(E,\omega\right) \omega^{2n} \right)=0.
\eea
Solving the equation, we find that the new saddle point is located at
\bea
\omega_{*} = 2E \sqrt{\frac{n}{\frac{3N-2}{2}+n}}. \label{firstsaddleeq}
\eea
\begin{figure}
\centering
\includegraphics[width=1\linewidth]{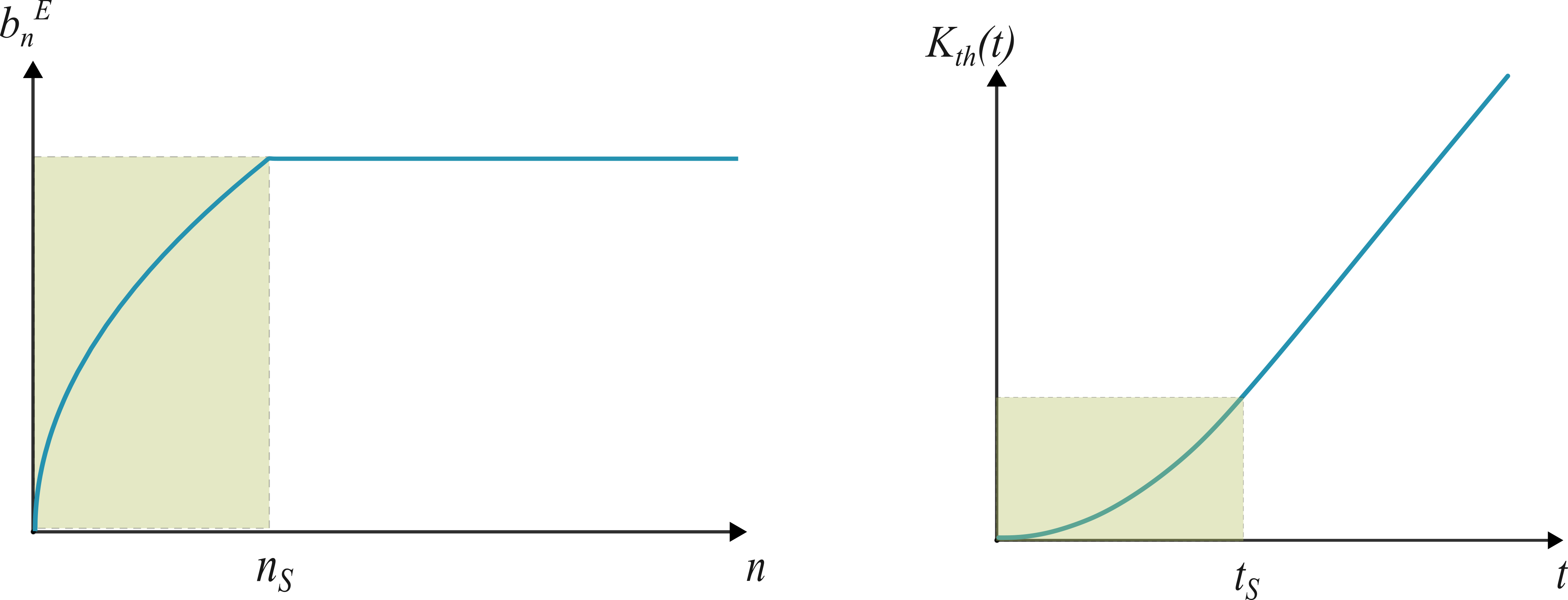}
\caption{The figure on the left shows the growth of the Lanczos coefficients w.r.t $n$. There is an initial scrambling phase (shaded in green) where $b_n^{E}$ grows as $\sqrt{n}$, and then it saturates to a constant value. The figure on the right shows the corresponding Krylov Complexity growth. During the scrambling phase, the K-complexity grows quadratically. Then it transitions into a linear growth proportional to the average energy of the box.}
\label{singleplotfig}
\end{figure}
\noindent When $n \gg \frac{3N L^2}{2\lambda^2}$,
\bea
\omega_{*} \simeq 2E .
\eea
It is easy to verify that \eqref{firstsaddleeq} satisfies \eqref{secondsaddleconsistanteq} when $n \gg \frac{3N L^2}{2\lambda^2}$. Therefore, the saddle point of the integral changes when we look at higher moments. This has interesting consequences for the behavior of Lanczos coefficients. In particular, we can use \eqref{momentlanczos} to see that:
\bea
\left\langle b_n^E \right\rangle_{\text{EE}} \sim 2E.
\eea
The saturation of Lanczos coefficients is referred to as ``Lanczos plateau''. This gives us
\bea
\left\langle b_n^E \right\rangle_{\text{EE}} \sim \begin{cases}\sqrt{\frac{8\lambda^2E^2n}{3L^2N}}, & n<\frac{3N L^2}{2\lambda^2} \\ E, & n>\frac{3N L^2}{2\lambda^2}\end{cases}
\eea
From \cite{Barbon:2019wsy,Fan:2022xaa,noh2021operator}, we can see that these Lanczos coefficients result in an initial \textit{scrambling phase} where the Krylov complexity grows quadratically. Following the scrambling phase, K-complexity switches to linear growth. Working out the growth rates and reinstating factors of $\hbar$, we find that the Krylov complexity is given by
\bea
\left\langle K^{E}(t) \right\rangle_{\text{EE}} \sim \begin{cases}\frac{8\lambda^2E^2}{3L^2N\hbar^2}t^2, & t<t_S\\ \frac{E}{\hbar}\ t, & t>t_S\end{cases} \label{microKeq}
\eea
where $t_S$ is the scrambling time. We can determine $t_S$ by noting that the dynamics of the operator can be thought of as a particle moving on a one-dimensional semi-infinite chain. The sites on the chain are labeled by $n$, and Krylov complexity \eqref{fixedekrylovdef} measures the average position $\langle n \rangle$. The saddle points change when $n \sim \frac{3N L^2}{2\lambda^2} \equiv n_S$. Therefore, scrambling time is the amount of time Krylov complexity takes to reach $n_S$. This gives us
\bea
t_S \simeq \frac{\beta L^2\hbar}{2\lambda^2} \ \text{where} \ \beta = \frac{1}{k_BT}
\eea
Using \eqref{thermalKrylovdef}, we can compute the thermal K-complexity in the eigenstate ensemble:
\bea
\left\langle K_{t h}\right\rangle_{\text{EE}} =\frac{\int_0^{\infty} d E e^{-\beta E} \mathcal{C}(E) \left\langle K^{E}(t) \right\rangle_{\text{EE}} }{\int_0^{\infty} d E e^{-\beta E} \mathcal{C}(E)}
\eea
It is easy to see that we get
\bea
\left\langle K_{t h}\right\rangle_{\text{EE}} \sim \begin{cases}\frac{8\lambda_{*}^2E_{*}^2}{3L^2N\hbar^2}t^2, & t<t_S\\ \frac{E_{*}}{\hbar}\ t, & t>t_S\end{cases} \label{microKeq}
\eea
where $E_{*}$ and $\lambda_{*}$ are the average energy and thermal wavelength of the box. Plotting these functions, we obtain figure \ref{singleplotfig}.
\subsection{Slowly Leaking Gas}
\begin{figure}
\centering
  \includegraphics[width=1\linewidth]{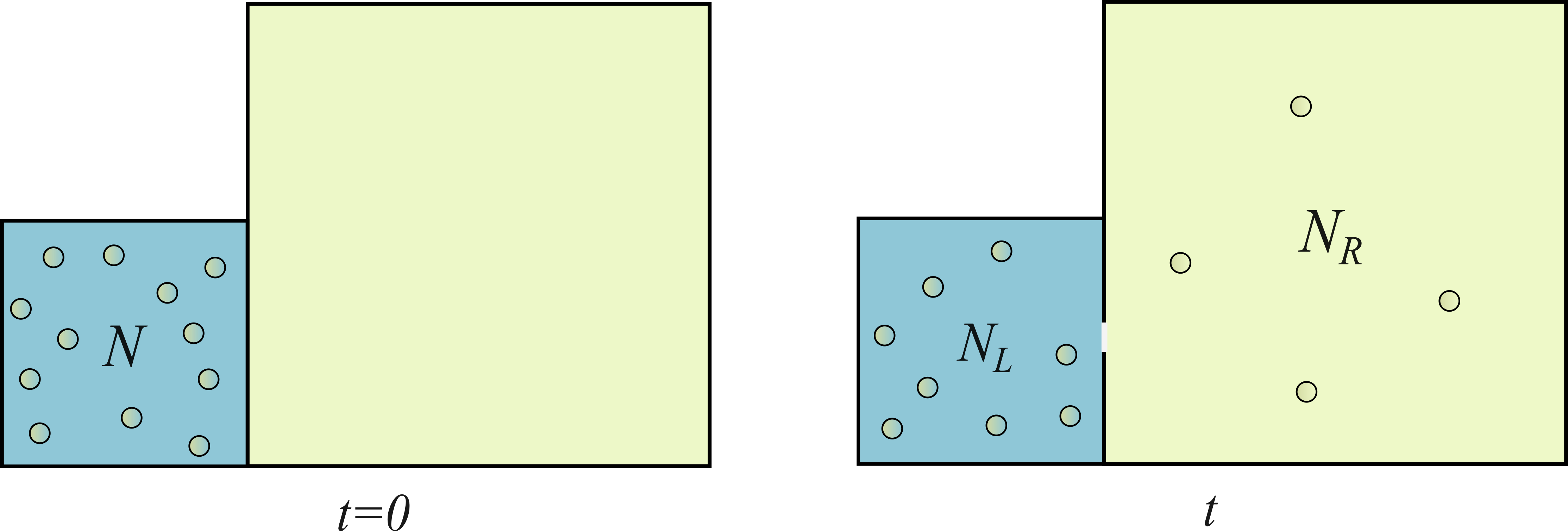}
\caption{Consider two cubic boxes in contact with each other. We fill the box on the left with $N$ hard spheres. Let us assume that we are in the semi-classical limit where we can localize particles. At time $t=0$, we poke a small hole on their common wall so that the gas leaks slowly into the right box. Let us denote the instantaneous number of particles in the boxes by $N_L$ and $N_R$.}
\label{leakinggasfig}
\end{figure}
Now let us look at the slowly leaking gas model used in \cite{Krishnan:2021faa}. Consider two cubic boxes sharing a common side as in figure \ref{leakinggasfig}. Let us assume that the left box has $N$ hard spheres while the right box is empty. At time $t=0$, we make a small hole in their shared wall so that the gas leaks slowly into the right box. We assume that the gas is leaking so slowly that there exists a time scale over which both the boxes have separately equilibrated. We will refer to this period as an \textit{epoch}. During each epoch, the number of particles in each box, denoted by $N_{L,R}$, remains roughly constant. Since we are working in the semiclassical limit, we can always use either $N_L$ or $N_R$ to characterize each epoch.
\subsubsection{Krylov Complexity during an epoch}
\label{Kcompepochsection}
Within an epoch, there is no net exchange of particles between the boxes. Consequently, when examining the full Hamiltonian, it becomes evident that during each epoch, the Hamiltonian takes on the following factorized structure:
\bea
H\simeq H_L \otimes \mathds{1}_R+ \mathds{1}_L \otimes H_R \label{factorizedH}
\eea
Consider an operator $P_{1}$, which measures the momentum of one of the particles in the left box. The operator has the following form
\bea
P_{1} = P_{1,L} \otimes \mathds{1}_R. \label{momentumfactoreq}
\eea
Here $P_{1,L}$ is an operator that acts only on the left box. Now let us look at the Krylov complexity of this operator. If we are computing the Krylov complexity w.r.t to the entire system, we will get the results of the previous section - a scrambling phase followed by linear growth. This is because the left and right boxes form a closed classically chaotic system, and we anticipate a version of Berry's conjecture to be applicable to the eigenstates of the entire system. However, we are after the Krylov complexity of the left box alone, which is an open quantum system.

The evolution of an operator $O_L$ acting on the left box is given by the master equation:
\bea
\dot{O_L}(t) = \frac{i}{\hbar}\left(\mathcal{L}_L+\mathcal{L_{D}}\right)
\eea
Here, $\mathcal{L}_L$ denotes the Liouvillian of the left box, and $\mathcal{L}_{D}$ represents a dissipative term arising from the interaction between the two boxes. The dissipative term makes the evolution non-Hermitian, and there is no consensus on extending Krylov complexity calculations to open quantum systems \cite{Bhattacharjee:2022lzy,Bhattacharya:2022gbz,Liu:2022god}. However, the ``effective'' Hamiltonian has no interaction term during an epoch. Therefore, the evolution of an operator $O_L$ would be controlled only by $\mathcal{L}_L$, or equivalently, the Hamiltonian of the left box. This allows us to carry over our definition of moments \eqref{momenteq} to the two-box system:
\bea
\left(\mu_{2 n}^{E}\right)_L = \frac{1}{\mathcal{C}(E_L)}\int_{-2E_L}^{2E_L} d\omega \ \rho_0(E_L, \omega_L) \left|\left\langle E_L+\frac{\omega_L}{2}\left|O_L\right| E_L-\frac{\omega_L}{2}\right\rangle\right|^2 \omega_L^{2n} \label{momenteqdefleft}
\eea
where $E_L\pm \frac{\omega_L}{2}$ are the energy eigenvalues of the left box. Now let us look at the operator $P_1$. The operator acting on the left box is given by tracing out the degrees of freedom of the right box. This gives us $\Tr_R{\left(P_{1}\right)}$. The moments of this operator are given by
\bea
\left(\mu_{2 n}^{E}\right)_L = \frac{1}{\mathcal{C}(E_L)}\int_{-2E_L}^{2E_L} d\omega \ \rho_0(E_L, \omega_L) \left|\left\langle E_L+\frac{\omega_L}{2}\left|\Tr_R{\left(P_{1}\right)}\right| E_L-\frac{\omega_L}{2}\right\rangle\right|^2 \omega_L^{2n} \label{momenteq2}
\eea
Using \eqref{momentumfactoreq}, we get
\bea
\left(\mu_{2 n}^{E}\right)_L = \frac{1}{\mathcal{C}(E_L)}\int_{-2E_L}^{2E_L} d\omega \ \rho_0(E_L, \omega_L) \left|\left\langle E_L+\frac{\omega_L}{2}\left|P_{1,L}\right| E_L-\frac{\omega_L}{2}\right\rangle\right|^2 \omega_L^{2n} \label{momenteq3}
\eea
which is precisely the same expression we had in \eqref{momenteq}. Therefore, the calculations in the previous section will go through - The K-complexity will grow linearly, following a quadratic scrambling phase.
\begin{figure}
\centering
  \includegraphics[width=0.75\linewidth]{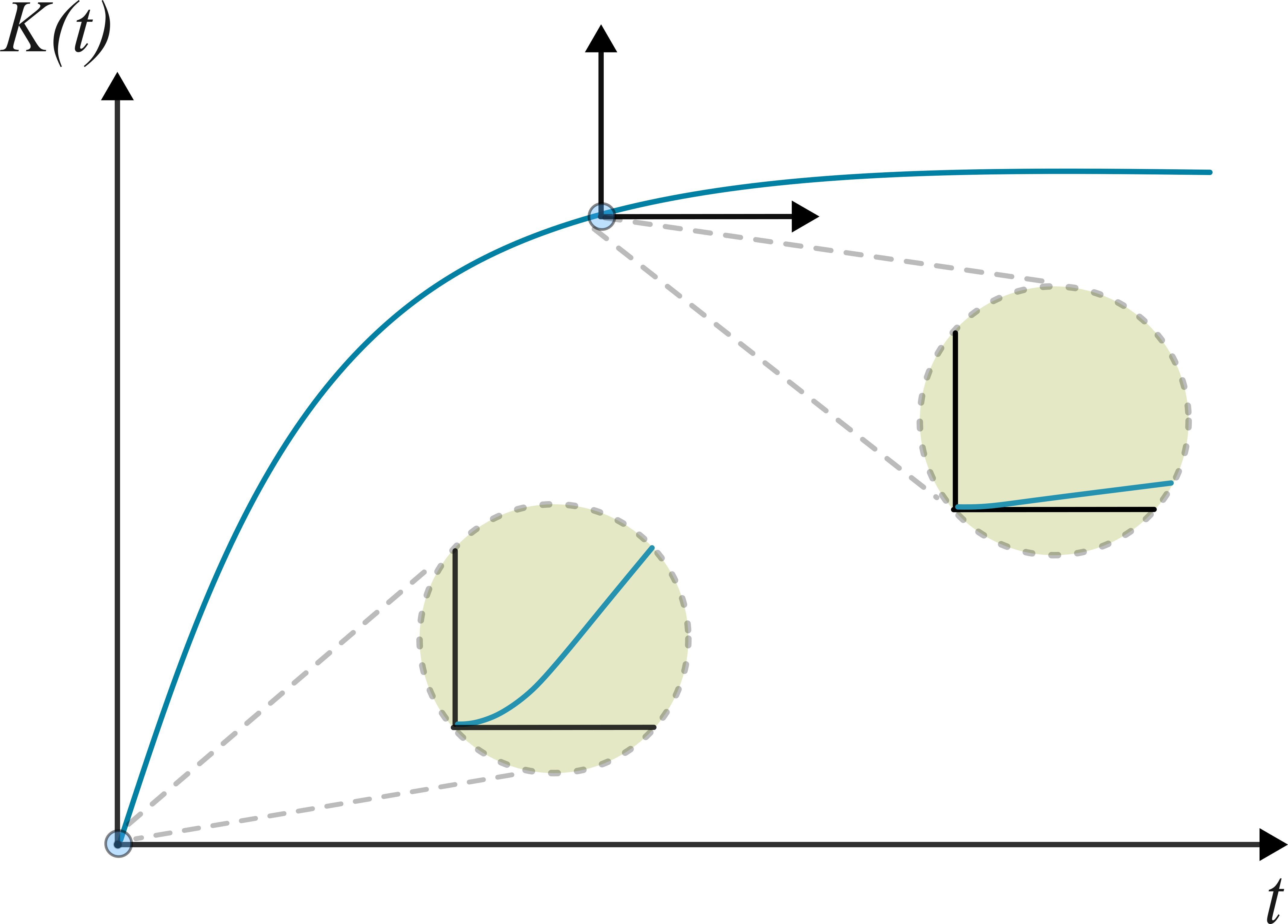}
\caption{The figure shows the thermal Krylov complexity of a slowly leaking gas as a function of time. The complexity continues to rise, but eventually levels off. The inset figures show how K-complexity grows during two different epochs. As we move into future epochs, we observe a decrease in the late time growth rate.}
\label{plotfullfig}
\end{figure}
\subsubsection{Stitching Together Epochs}
From \eqref{microKeq}, we can see that the late time linear growth rate of Krylov complexity during an epoch is given by the average energy of the left box, which we will denote by $E_L$. Using $E = \frac{3}{2}Nk_BT$, we get
\bea
\frac{dK_{th}(t)}{dt} \sim \frac{E_L}{\hbar} =\frac{3}{2\hbar} N_Lk_BT \label{epochKcompeq}
\eea
Now, we will compute the time dependence of $N_L$. Suppose the right box is sufficiently large for the gas to leak out completely. If the area of the hole is given by $A$, the leakage rate is given by \cite{Schroederbook}
\bea
\frac{d N}{d t}=-\frac{A}{2 L^3} \sqrt{\frac{k T}{m}} N .
\eea
We can immediately integrate the above equation if we assume that the temperature of the left box remains constant. This gives us
\bea
\begin{aligned}
N(t)& = N e^{-\frac{A}{2 L^3} \sqrt{\frac{k T}{m}}t}\\
&\equiv N e^{-\frac{t}{t_L}}
\end{aligned}
\eea
where $t_L$ is the leakage time of the system. Treating $N(t)$ as an instantaneous value, we can integrate \eqref{epochKcompeq} to obtain
\bea
K_{th}(t) \sim \frac{N\sqrt{mkT}L^3}{A\hbar}\left(1-e^{-\frac{t}{t_L}}\right)
\eea
Plotting the above function, we get figure \ref{plotfullfig}. Krylov complexity keeps increasing and eventually levels off when $t=t_L$.

\section{Holographic Complexity of an Evaporating Black Hole}
\label{cvsection}
In this section, we will study the holographic complexity of a slowly evaporating black hole using the Complexity=Volume prescription \cite{Susskind:2014rva,Stanford:2014jda}. To make the calculations tractable, let us model the black hole by patching together a sequence of $k$ static Schwarzschild spacetimes across negative energy null shock waves. The metric of the $d+1$ dimensional black hole is then given by
\bea
ds^2 = -F(r,v)dv^2+2dvdr+r^2d\Omega_{d-1}
\eea
where
\bea
F(r,v) = 1-\frac{f(v)}{r^{d-2}}.
\eea
We will choose $f$ to have the following profile
\bea
f(v) = \begin{cases}\omega_1^{d-2}, & v<v_1\\
\cdots\\
\omega_i^{d-2}, & v_{i-1}<v<v_{i}\\
\cdots\\
0, & v_{k-1}<v<v_{k}\end{cases}
\eea
The mass of each patch is given by
\bea
M_i=\frac{(d-1) \Omega_{d-1}}{16 \pi G_N}\omega_i^{d-2},
\eea
where
\bea
M_1>M_2 > \cdots>0.
\eea
For the patched-up spacetime to be a good approximation to an evaporating black hole, we will assume the width of each patch to be much smaller than the time scales at which the black hole mass changes considerably. Moreover, we will also assume that the width is larger than the scrambling time of the black hole. Each patch corresponds to a period where the black hole is effectively not evaporating. Therefore, we will adopt the terminology from our previous section and refer to each patch as an epoch. In particular, the patch between $v=v_{i-1}$ and $v=v_{i}$ shock waves will be labeled as the $i$-th epoch. During an epoch, the black hole has a constant mass, and we can rewrite $F(r,v)$ as $F_i(r)$.
\subsection{Penrose Diagram}
\label{penrosesection}
\begin{figure}
\centering
  \includegraphics[width=0.6\linewidth]{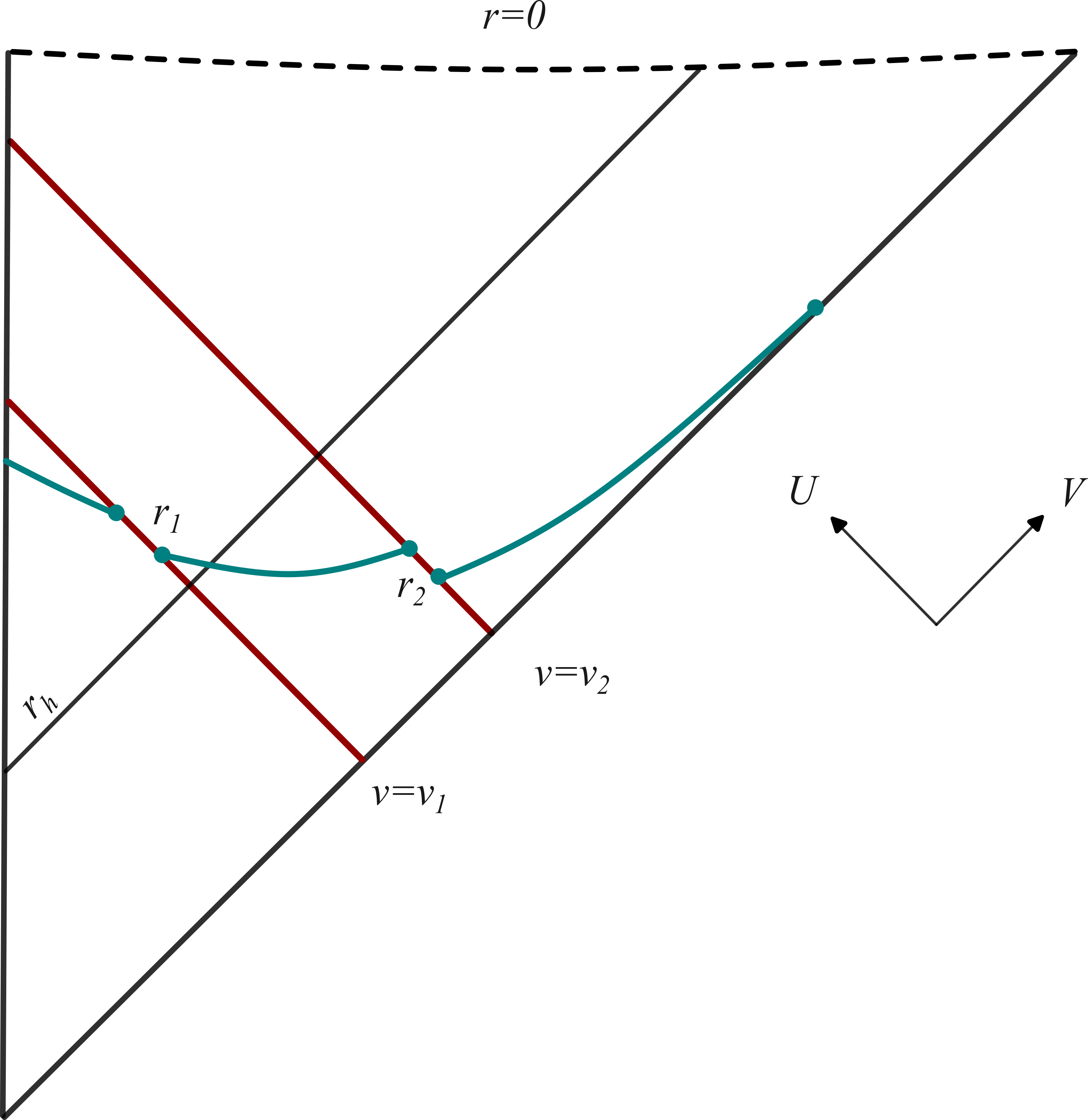}
\caption{Penrose diagram of the black hole spacetime when there are three epochs. We will assume the mass of each epoch to satisfy the relation $M_1>M_2>M_3 \neq 0$. The horizon is indicated by the 45-degree line. The shock waves are marked by the red lines. A spacelike surface, indicated by the teal color, will be disconnected in the diagram. The end points of these disconnected surfaces follow the ordering \eqref{orderingeq}.}
\label{penrosediagfig0}
\end{figure}
To understand the structure of the spacetime, it is instructive to draw its Penrose diagram. To simplify the discussion, we will restrict ourselves to $3+1$-dimensions in this subsection. During an epoch, the tortoise coordinate is given by
\bea
r^{*}_i(r) = \int^r \frac{dr^{\prime}}{F_i(r^{\prime})} = r+2G_NM_i\log\left(\left|\frac{r}{2G_NM_i}-1\right|\right)
\eea
The corresponding outgoing Eddington–Finkelstein coordinates $u_i$ are defined as follows
\bea
u_i \equiv v - 2 r^{*}_i(r)\label{eddingtoneq}
\eea
It is easy to see that $u$ is discontinuous across the boundaries of the epochs. Therefore, if we go along a continuous curve, the coordinate $u$ has a ``jump'' in its value as soon as we cross a shock wave. Consequently, employing identical $u$ and $v$ coordinates throughout all epochs would render continuous curves discontinuous in the corresponding Penrose diagram. This discontinuity is the cost we have to pay to keep the Penrose undeformed \cite{Chapman:2018dem,Chapman:2018lsv}.

During an epoch, the outgoing null Kruskal coordinate $U_i$ can be defined as follows \cite{Misner:1973prb}:
\bea
U_i =\begin{cases}
-e^{-\frac{u_i}{4G_NM}}, & \text{Outside the horizon}\\
e^{-\frac{u_i}{4G_NM}}, & \text{Inside the horizon}\end{cases} \label{kruskalu}
\eea
Since $v$ is globally defined, we can define the outgoing null Kruskal coordinate $V$ everywhere as
\bea
V = e^{\frac{v}{4G_NM}}.
\eea
Using the definition of Kruskal coordinates, we can see that the horizon of the epochs will always be at $U_i=0$. Now let us figure out what happens to a continuous curve as it crosses a shock wave. We will denote the radial coordinate of the surface at the location of the shock wave by $r_S$. Using \eqref{eddingtoneq} and \eqref{kruskalu}, it is easy to verify that when $r_S$ is either in the interior or sufficiently far away from the horizon, we have the relation
\bea
U_{i}(r_S)>U_{i+1}(r_S) \label{orderingeq}
\eea
Now let us draw the Penrose diagram of the spacetime. We will use the same $U$ and $V$ coordinates across all the epochs. This gives us figure \ref{penrosediagfig0}. A connected surface will be discontinuous in this diagram. We can locate the end points of these disconnected pieces by using \eqref{orderingeq}.

\subsection{Complexity=Volume}
We can study the growth of complexity using the Complexity=Volume conjecture \cite{Stanford:2014jda,Susskind:2014rva}. Let us assume that there are only three epochs, with masses satisfying the relation $M_1>M_2>M_3 \neq 0$. We will see that extending our results to an arbitrary number of epochs is straightforward. The Penrose diagram of this spacetime is given in figure \ref{penrosediagfig0}.

Consider spherically symmetric spacelike codimension-1 surfaces anchored onto a cutoff surface in the asymptotic region. We will assume that the cutoff surface is at $r=r_{\infty}$ and the boundary anchoring time is denoted by $t$. The volume of this surface is given by
\bea
\mathcal{V}=\Omega_{d-1} \int d \lambda r^{d-1} \sqrt{-F(r,v) \dot{v}^2+2 \dot{v} \dot{r}} \label{volumeeq}
\eea
Here $\Omega_{d-1}$ is the dimensionless area of the $(d-1)$-dimensional unit sphere. Let us rewrite the volume integral as the following summation
\bea
\mathcal{V} = \sum_{i} \mathcal{V}_i
\eea
where $\mathcal{V}_i$ is the volume of the portion of the surface located within the $i$-th epoch. We can then express each of these terms as follows:
\bea
\mathcal{V}_i=\Omega_{d-1} \int d \lambda r^{d-1} \sqrt{-F_i(r) \dot{v}^2+2 \dot{v} \dot{r}} \equiv \Omega_{d-1} \int d \lambda \mathcal{L}_i
\eea
We can see that $\mathcal{L}_i$ is independent of $v$. Therefore, there is a conserved quantity associated with the integral, which we will denote by $E_i$:
\bea
E_i=\frac{\partial \mathcal{L}_i}{\partial \dot{v}}=\frac{r^{d-1}(\dot{r}-F_i\dot{v})}{\sqrt{-F_i \dot{v}^2+2 \dot{v} \dot{r}}} \label{energyeqextre}
\eea
The volume integral is reparametrization invariant, allowing us to choose
\bea
r^{d-1} \sqrt{-F_i \dot{v}^2+2 \dot{v} \dot{r}}=1 \label{paraeqextre}
\eea
The maximal volumes are given by extremizing the action in \eqref{volumeeq}. This gives us the following equations of motion:
\bea
\begin{aligned}
E_i &=r^{2(d-1)}(\dot{r}-F_i(r) \dot{v}) \\
r^{2(d-1)} \dot{r}^2 &=F_i(r)+r^{-2(d-1)} E_i^2
\end{aligned} \label{paraeqextre1}
\eea
The part of the surface within an epoch can have a \textit{turning point} where $\dot{r}$ vanishes. We will denote this point by $r_\text{i,min}$. We will see later that it is \textit{not} necessary for the surface to have a turning point in an epoch. However, these turning points will be crucial in calculating late time growth rates during an epoch.

To characterize various features of the extremal volume surfaces, it is convenient to define an effective potential as follows:
\bea
V_{i}(r) =F_i(r)r^{2(d-1)}+ E_i^2. \label{effpotentialvolumeeq1}
\eea
In particular, we can obtain the turning point $r_\text{i,min}$ from the zero of the effective potential
\bea
V_{i}(r_\text{i,min}) = 0 \quad \implies \quad F_i(r_\text{i,min})r_\text{i,min}^{2(d-1)}+ E_i^2=0\label{turningpointenergyrelation}
\eea
Using the equations of motion, we can rewrite the volume integral as follows:
\bea
\mathcal{V}= \Omega_{d-1} \sum_i\int d r \frac{r^{2(d-1)}}{\sqrt{F_i(r) r^{2(d-1)}+E_i^2}}\label{volrintegral}
\eea

The Complexity=Volume proposal suggests that the complexity of the black hole at time $t$ is given by the volume of these extremal surfaces \cite{Stanford:2014jda}:
\bea
\mathcal{C}_i= \frac{\mathcal{V}(t)}{G_N\omega_i}
\eea
where $\omega_i$ is the horizon radius during the $i$-th epoch.

\begin{figure}
\centering
\includegraphics[width=1\linewidth]{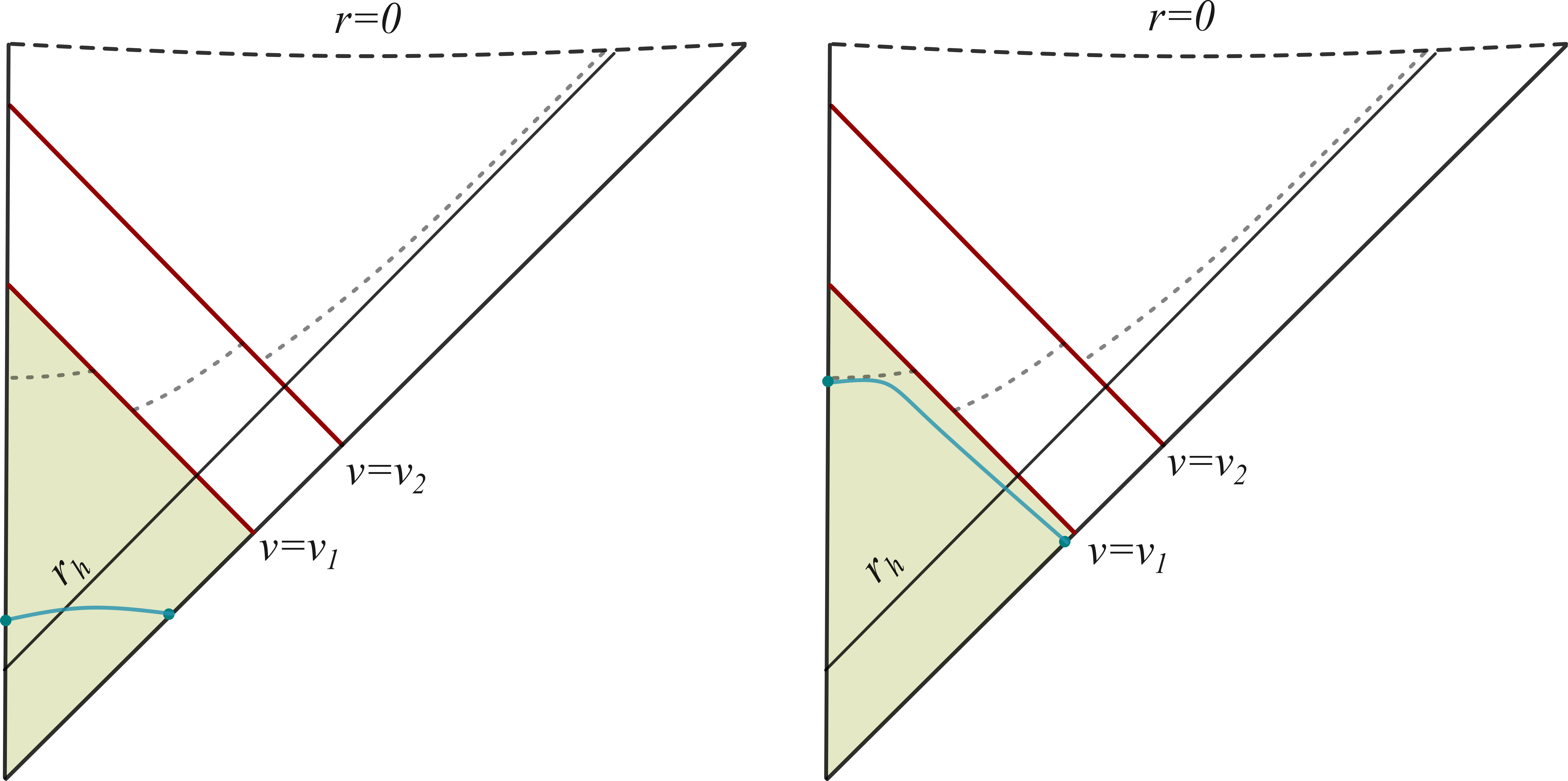}
\caption{The left (right) figure shows the early (left) time behavior of the extremal volume surface when the boundary anchoring points are in the first epoch. We have not included the cutoff surface in the diagrams to avoid cluttering. The 45-degree line depicts the horizon. The grey dashed lines in the interior of the black hole correspond to the accumulation surfaces (Ref \eqref{accumulationeqpoch1}) of the epochs.}
\label{penrosepoch1fig}
\end{figure}

\begin{figure}
\centering
\includegraphics[width=1\linewidth]{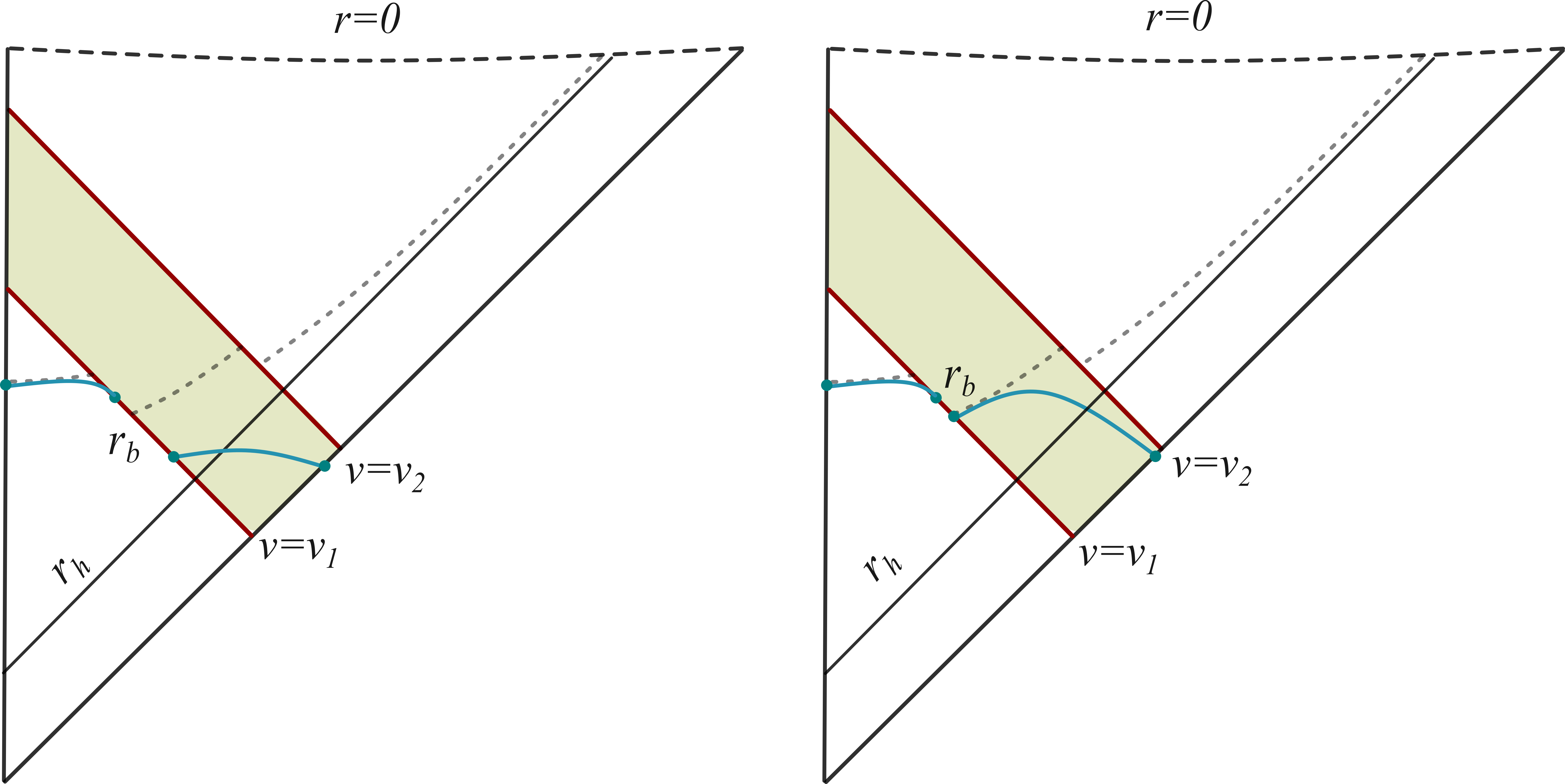}
\caption{The left (right) figure shows the early (left) time behavior of the extremal volume surface when the boundary anchoring points are in the second epoch. The extremal surface is disconnected, and the endpoints of the surface are located using the relation \eqref{orderingeq}.}
\label{penroseepoch2fig}
\end{figure}

\textbf{First Epoch}: Now let us look at the case where the boundary anchoring point is in the first epoch (Refer figure \ref{penrosepoch1fig}). This case reduces to the calculation of extremal volumes in a single-sided Schwarzschild black hole. Since we are in the first epoch, there will only be one term in the sum \eqref{volrintegral}. The surface will always have a turning point. At late times, $r_\text{1,min}$ approaches a constant radial surface in the interior of the black hole, located at the critical point of effective potential \cite{Stanford:2014jda, Carmi:2017jqz}. We will refer to this surface as the \textit{accumulation surface}. The location of this surface is given by
\bea
V^{\prime}_{1}(R_\text{1,min}) =0 \implies \ R_\text{1,min}=\omega_1\left(\frac{d}{2d-2}\right)^{\frac{1}{d-2}}. \label{accumulationeqpoch1}
\eea
Since the width of each epoch is much larger than the scrambling time, the turning points will necessarily approach the accumulation surface at late times. Now let us calculate the growth rate of \eqref{volumeeq} as a function of the boundary anchoring time. From \eqref{paraeqextre1}, we can see that
\bea
t+r_{\infty}^*-r^*\left(r_{1,\min}\right)=\int_{v_{1,\min }}^{v_{\infty}} d v=\int_{r_{1,\min }}^{r_\infty} d r\left[\frac{-E_1}{F_1(r) \sqrt{F_1(r) r^{2(d-1)}+E_1^2}}+\frac{1}{F_1(r)}\right]\label{timevoleq}\nonumber\\
\eea
Using \eqref{timevoleq}, we can rewrite the volume integral \eqref{volrintegral} as follows:
\bea
\frac{\mathcal{V}}{ \Omega_{d-1}}=\int_{r_{1,\min }}^{r_{\max }} d r\left[\frac{\sqrt{F_1(r) r^{2(d-1)}+E_1^2}}{F_1(r)}-\frac{E_1}{F_1(r)}\right]+E_1\left(t+r_{\infty}^*-r^*\left(r_{1,\min }\right)\right)
\eea
Taking a derivative w.r.t $t$ and using Leibniz integral rule, we find the simple relation
\bea
\begin{aligned}
\frac{d\mathcal{V}}{dt} &= \Omega_{d-1}E_1.\\
& = \Omega_{d-1}\sqrt{-F_{1}(r_{1,\min})}r_{1,\min}^{d-1}
\end{aligned}\label{volgrowtheq1}
\eea
where we have used \eqref{turningpointenergyrelation} to rewrite the energy in terms of the metric components. At late times, $r_{1,\min}$ approaches $R_{1,\min}$, a constant. Using \eqref{accumulationeqpoch1}, we find that
\bea
\frac{d\mathcal{V}}{dt} =c_d \Omega_{d-1} \omega_1^{d-1}
\eea
where
\bea
c_d = \sqrt{\frac{d-2}{d}} \left(\frac{d}{2d-2}\right)^{\frac{d-1}{d-2}}\label{constantfactoreq}
\eea
Therefore, the late time complexity growth during the first epoch is given by
\bea
\frac{d\mathcal{C}_1}{dt} = \frac{1}{G_N\omega_1}\frac{d\mathcal{V}}{dt} = \frac{16\pi c_d}{d-1}M_1
\eea
\textbf{Second epoch}: Now let us look at the case where the boundary anchoring point is in the second epoch (see figure \ref{penroseepoch2fig}).
Following the discussion in section \ref{penrosesection}, the extremal volume surfaces will be disconnected in the Penrose diagram. The endpoints satisfy the ordering relation \eqref{orderingeq}. We will label the radial coordinate of the point where the extremal surface intersects the epoch's boundary by $r_S$. The volume functional \eqref{volrintegral} is the sum of two terms:
\bea
\mathcal{V}= \Omega_{d-1} \int_{r_{1,\min }}^{r_S} d r \frac{r^{2(d-1)}}{\sqrt{F_1(r) r^{2(d-1)}+E_1^2}}+ \Omega_{d-1} \int_{r_{S}}^{r_\infty} d r \frac{r^{2(d-1)}}{\sqrt{F_2(r) r^{2(d-1)}+E_2^2}}
\eea
The surface will always have a turning point in the first epoch. Moreover, the turning point will be at the accumulation surface of the first epoch $R_\text{1,min}$. Three possibilities arise when we consider the turning point of the second epoch. The surface will not have a turning point during the very early stages. However, as boundary anchoring time increases, the surface will develop a turning point in the interior of the black hole. At late times, the turning point $r_{2,\min}$ will approach the accumulation surface of the epoch, given by
\bea
V^{\prime}_{2}(R_\text{2,min}) =0 \implies \ R_\text{2,min}=\omega_2\left(\frac{d}{2d-2}\right)^{\frac{1}{d-2}}. \label{accumulationeqpoch2}
\eea

Now let us calculate the growth rate of these volumes. As in the previous section, we have
\bea
v_1-r^*\left(r_{1,\min}\right)=\int_{v_{1,\min }}^{v_{1}} d v=\int_{r_{1,\min }}^{r_S} d r\left[\frac{-E_1}{F_1(r) \sqrt{F_1(r) r^{2(d-1)}+E_1^2}}+\frac{1}{F_1(r)}\right]\label{timevoleq2}
\eea
and
\bea
t+r_{\infty}^*-v_1=\int_{v_{1}}^{v_{\infty}} d v=\int_{r_{S}}^{r_\infty} d r\left[\frac{-E_2}{F_2(r) \sqrt{F_2(r) r^{2(d-1)}+E_2^2}}+\frac{1}{F_2(r)}\right]\label{timevoleq3}
\eea
This gives us
\bea
\begin{aligned}
\frac{\mathcal{V}}{ \Omega_{d-1}}&=\int_{r_{1,\min }}^{r_{S}} d r\left[\frac{\sqrt{F_1(r) r^{2(d-1)}+E_1^2}}{F_1(r)}-\frac{E_1}{F_1(r)}\right]+E_1\left(v_1-r^*\left(r_{1,\min}\right)\right)\\
&+\int_{r_{S}}^{r_{\max }} d r\left[\frac{\sqrt{F_2(r) r^{2(d-1)}+E_2^2}}{F_2(r)}-\frac{E_2}{F_2(r)}\right]+E_2\left(t+r_{\infty}^*-v_1\right)
\end{aligned}
\eea
Taking a time derivative w.r.t $t$, we get
\bea
\begin{aligned}
\frac{1}{\Omega_{d-1}}\frac{d\mathcal{V}}{ dt}&=E_2 \\
&+\frac{dr_S}{dt}\left[\frac{\sqrt{F_1(r) r^{2(d-1)}+E_1^2}}{F_1(r)}-\frac{E_1}{F_1(r)}-\frac{\sqrt{F_2(r) r^{2(d-1)}+E_2^2}}{F_2(r)}-\frac{E_2}{F_2(r)}\right]_{r=r_S}
\end{aligned}
\eea
From \eqref{paraeqextre1}, it is easy to see that the term in the second line is proportional to $\dot{v}_1(r_S)-\dot{v}_2(r_S)$. Since $v$ is continuous across each epoch, this function vanishes, and we get the simple expression:
\bea
\begin{aligned}
\frac{d\mathcal{V}}{ dt}&=\Omega_{d-1} E_2 = \Omega_{d-1}\sqrt{-F_{2}(r_{2,\min})}r_{2,\min}^{d-1}
\end{aligned}
\eea
At late times, $r_{2,\min}$ approaches the accumulation surface \eqref{accumulationeqpoch2}. Substituting \eqref{accumulationeqpoch2} in the above equation, we find the late time complexity growth during the second epoch to be given by
\bea
\begin{aligned}
\frac{d\mathcal{C}_2}{dt} &= \frac{1}{G_N\omega_2}\frac{d\mathcal{V}}{dt}\\
&\simeq \frac{\Omega_{d-1}}{G_N\omega_2}\sqrt{-F_{2}(R_{2,\min})}R_{2,\min}^{d-1}\\
& = \frac{\Omega_{d-1}}{G_N}\sqrt{\frac{d-2}{d}} \left(\frac{d}{2d-2}\right)^{\frac{d-1}{d-2}} \omega_2^{d-2}\\
&=\frac{16\pi c_d}{d-1}M_2,
\end{aligned}
\eea
where $c_d$ is defined in \eqref{constantfactoreq}.

\noindent \textbf{i-th epoch}: It is easy to extend the results to the $i$-th epoch. Complexity will undergo an initial transitional phase, after which it will settle to a linear growth characterized by the growth rate:
\bea
\frac{d\mathcal{C}_i}{dt} \simeq \frac{16\pi c_d}{d-1}M_i \label{epochgrowthrateCV}
\eea
Now, let us look at the overall trend of complexity growth. We will do this by taking a ``continuum'' limit where the width of each epoch goes to zero. Note that we cannot formally take this limit since the scale separation between the scrambling time and the width of an epoch was crucial for our calculation to be valid. One should think of the limit as a means to smooth out the otherwise jagged curve that would result from combining complexity across various epochs. Operationally, we proceed by replacing $M(i)$ by the instantaneous mass of the black hole $M(t
)$. Let us assume that we are $3+1$-dimensions. If the black hole is a perfect black body that satisfies the Stefan-Boltzmann law, then we have \cite{Santi:2019lvi}
\bea
M(t) = M_0 \left(1-\frac{t}{t_E}\right)^{1/3}
\eea
where $t_E=\frac{5120\pi G_N^2M_0^3}{\hbar c^4}$ is the lifetime of the black hole.
\begin{figure}
\centering
\includegraphics[width=0.6\linewidth]{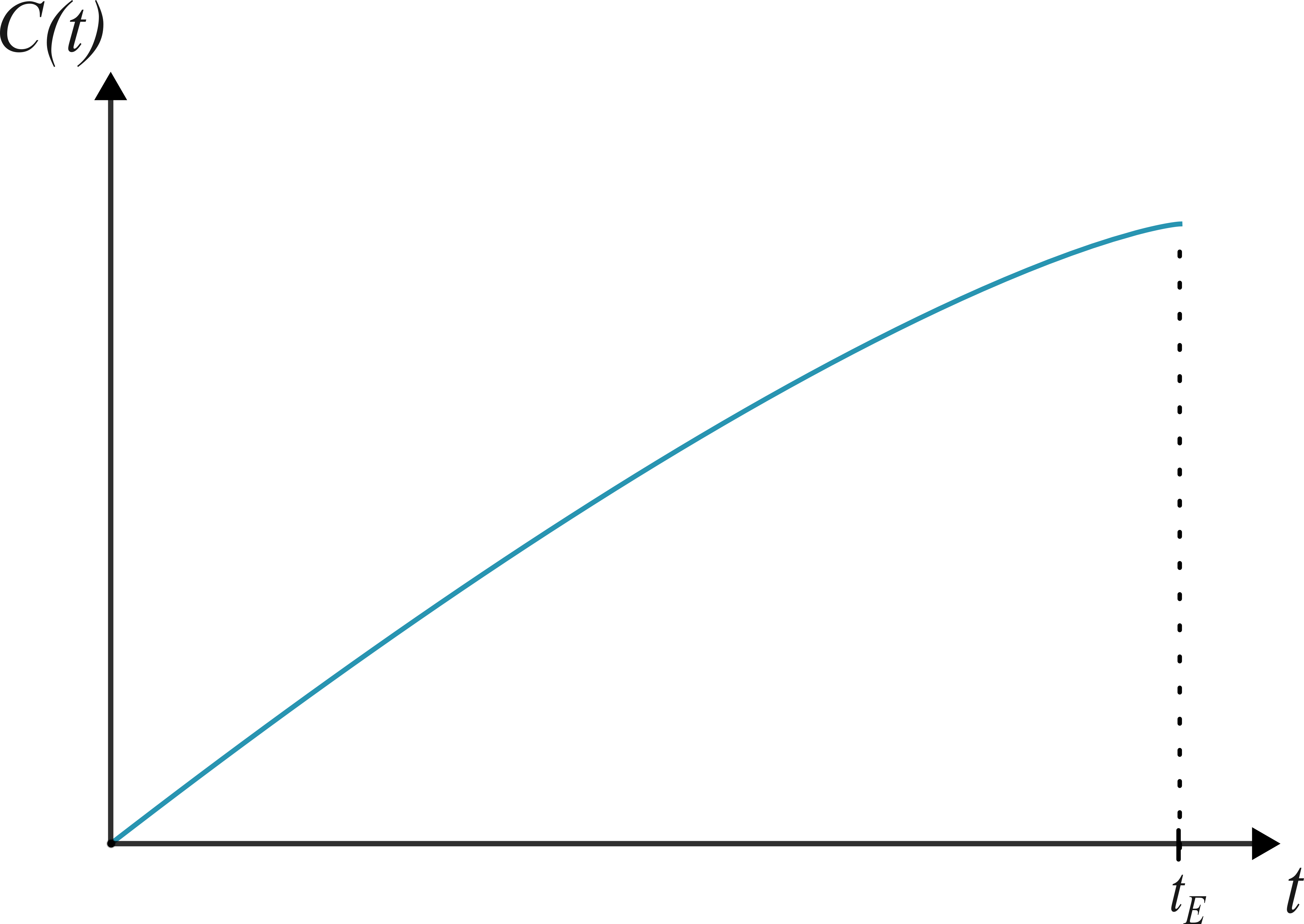}
\caption{Holographic complexity of an evaporating black hole obtained by taking the continuum limit of equation \eqref{epochgrowthrateCV}. Here $t_E$ is the lifetime of the black hole.}
\label{holoplotfullfig}
\end{figure}
Integrating the expression, we get
\bea
\boxed{C(t) = 2\sqrt{3}\pi M_0t_E\left(1-\left(1-\frac{t}{t_E}\right)^{4/3}\right)}\label{finalholoeq}
\eea
Plotting this function, we get figure \ref{holoplotfullfig}.
\section{Discussion}
Let us briefly review the calculations in section \ref{Kcompsection}. In conjunction with Berry's conjecture, the slow leakage assumption played a crucial role in rendering the Krylov complexity calculation analytically tractable. At first glance, the former assumption might appear limiting. However, slow leakage is \textit{required} if one wants to use intensive quantities like temperature at every instant of the process. The appearance of an ensemble-averaged semiclassical description in the context of black holes can be attributed to the existence of such quasi-static equilibriums \cite{Krishnan:2021faa}.

Working out details, we saw that the Krylov complexity had the following behavior:
\begin{itemize}
\item During an epoch, complexity goes through a scrambling phase and then transitions to linear growth.
\item Complexity keeps increasing even as we cross the boundary of each epoch. However, the late time linear growth rate decreases with each successive epoch.
\item When the gas has completely leaked out of the box, complexity levels off.
\end{itemize}
We claim that these results carry over to any chaotic quasi-static open quantum system if the operator under consideration satisfies ETH. Let us briefly outline this calculation by examining a system interacting with its environment. Consider an operator $O$ which satisfies the ETH ansatz. The off-diagonal elements of this operator are given by
\bea
\left\langle E_i\left|O\right| E_j\right\rangle \approx F\left(E,\omega\right)R_{ij}
\eea
where $E_{i,j}$ are the energy eigenstates of the system. $R_{ij}$ is a zero mean, unit variance random matrix whereas $E$ and $\omega$ are given by \eqref{energyaverageeq}. During an epoch, the Krylov complexity can be calculated using the Liouvillian of the system, following the assumptions in \ref{Kcompepochsection}. If we assume $F$ to decay as $\omega \to \infty$, then we can use the arguments in section 3.1 of \cite{Barbon:2019wsy} to see that the Krylov complexity undergoes a scrambling phase, followed by linear growth. As we have observed in the case of the slowly leaking gas, the linear growth rate will be proportional to the degrees of freedom of the system. Therefore, complexity will increase as we go from one epoch to the other, and it will eventually level off if all the degrees of freedom leak out of the system. This reproduces the advertised behavior.

The holographic complexity, computed using a gravity calculation, displays the same behavior we described earlier. We can push this parallel further by comparing the late time growth rates \eqref{epochKcompeq} and \eqref{epochgrowthrateCV}. From the laws of black hole thermodynamics, we can see that the mass of the black hole plays the role of energy \cite{Wald:1999vt}. Therefore, both calculations result in the same late time linear growth, provided we correctly identify the thermodynamic quantities on the black hole side. This provides further evidence to the claim that black holes can be described by a chaotic open quantum mechanical system with finite degrees of freedom when observed from the outside \cite{Susskind:1993if,Almheiri:2020cfm}. Another manifestation of this proposal can be found in \cite{Krishnan:2021faa}, where the entanglement entropy of the slowly leaking gas model matched the gravitational path integral calculation in \cite{Penington:2019kki}. We can put these statements on a firmer footing by thinking of it as a consequence of black hole complementarity \cite{Susskind:1993if}, which posits that the interior of a black hole can be thought to be described by a finite number of quantum mechanical degrees of freedom living on the stretched horizon of the black hole. The hard spheres in the left box assume the role of these degrees of freedom, while the particles in the right box model the outgoing Hawking radiation, allowing us to make a one-to-one map between an evaporating black hole and our two-box system.

Returning to the growth rate \eqref{epochgrowthrateCV}, we find that
\bea
\frac{dC}{dt} \sim M \propto ST \label{Susskindgrowtheq}
\eea
where $S$ and $T$ are the entropy and temperature of the black hole during that epoch. Our results are in tandem with \cite{Susskind:2018pmk} and the 2d gravity calculation performed in \cite{Schneiderbauer:2019anh}.

One can also calculate complexity using the Complexity=Action (CA) prescription \cite{Brown:2015bva,Brown:2015lvg,Schneiderbauer:2020isp}. However, to obtain meaningful results, the inclusion of a counterterm for null boundaries, similar to the one used in \cite{Chapman:2018dem,Chapman:2018lsv,Akhavan:2019zax,Omidi:2020oit}, might be necessary.
\acknowledgments
I thank Friðrik Freyr Gautason, Chethan Krishnan, Watse Sybesma, and Lárus Thorlacius for their insights and helpful discussions. This work was supported by the Icelandic Research Fund grant 228952-052.
\appendix

\bibliographystyle{JHEP}
\bibliography{refs}

\end{document}